\documentclass[journal=jacsat,manuscript=article]{achemso}

\usepackage[version=3]{mhchem} % Formula subscripts using \ce{}

\author{Chen Shen}
\affiliation{Institut f\"ur Materialwissenschaft, Technische Universit\"at Darmstadt, 64287, Darmstadt, Germany}

\author{Lei Wang}
\affiliation{Shenyang National Laboratory for Materials Science, Institute of Metal Research, Chinese Academy of Sciences, Shenyang 110016, P. R. China.}
\altaffiliation{School of Materials Science and Engineering, University of Science and Technology of China, Shenyang 110016, P. R. China.}

\author{Donghai Wei}
\affiliation{State Key Laboratory of Advanced Design and Manufacturing for Vehicle Body, College of Mechanical and Vehicle Engineering, Hunan University, 410082, Changsha, P. R. China}

\author{Yixuan Zhang}
\affiliation{Institut f\"ur Materialwissenschaft, Technische Universit\"at Darmstadt, 64287, Darmstadt, Germany}

\author{Guangzhao Qin}
\affiliation{State Key Laboratory of Advanced Design and Manufacturing for Vehicle Body, College of Mechanical and Vehicle Engineering, Hunan University, 410082, Changsha, P. R. China}
\email{gzqin@hnu.edu.cn}

\author{Xing-Qiu Chen}
\affiliation{Shenyang National Laboratory for Materials Science, Institute of Metal Research, Chinese Academy of Sciences, Shenyang 110016, P. R. China.}
\altaffiliation{School of Materials Science and Engineering, University of Science and Technology of China, Shenyang 110016, P. R. China.}
\email{xingqiu.chen@imr.ac.cn}

\author{Hongbin Zhang}
\affiliation{Institut f\"ur Materialwissenschaft, Technische Universit\"at Darmstadt, 64287, Darmstadt, Germany}
\email{hzhang@tmm.tu-darmstadt.de}

\title[An \textsf{achemso} demo]
  {Novel Two-Dimensional Layered \emph{M}Si$_2$N$_4$ (\emph{M} = Mo, W): New Promising Thermal Management Materials}

\abbreviations{IR,NMR,UV}

\begin{document}


\begin{abstract}
With the miniaturization and integration of nanoelectronic devices, efficient heat removal becomes a key factor affecting the reliable operation of the nanoelectronic device. With the high intrinsic thermal conductivity, good mechanical flexibility, and precisely controlled growth, two-dimensional (2D) materials are widely accepted as ideal candidates for thermal management materials.
In this work, by solving the phonon Boltzmann transport equation (BTE) based on first-principles calculations, we comprehensively investigated the thermal conductivity of novel 2D layered \emph{M}Si$_2$N$_4$ (\emph{M} = Mo, W).
Our results point to competitive thermal conductivities (162 W/mK) of monolayer MoSi$_2$N$_4$,
which is around two times larger than that of WSi$_2$N$_4$ and seven times larger than that of silicene despite their similar non-planar structures.
It is revealed that the high thermal conductivity arises mainly from its large group velocity and low anharmonicity. Our result suggests that MoSi$_2$N$_4$ could be a potential candidate for 2D thermal management materials.
\end{abstract}

\section{Introduction}
MoSi$_2$N$_4$ was synthesized by chemical vapor deposition (CVD) method \cite{hong2020chemical}.
It can be structural viewed as 2$H$-MoS$_2$-type MoN$_2$ intercalating into $\alpha$-InSe-type Si$_2$N$_2$ \cite{wang2020structure}.
The 2D MoSi$_2$N$_4$ was reported to exhibit semiconducting behavior, high carrier mobility, high strength, and excellent ambient stability\cite{hong2020chemical,wang2020structure}. In addition to that, due to its noncentrosymmetric hexagonal structure and unique electronic structure, several new physical properties such as second harmonic generation \cite{kang2020second}, valley pseudospin \cite{li2020valley,yang2021valley} and piezoelectricity \cite{guo2020intrinsic} were proposed in this system. However, at present, these researches mainly focus on the electronic properties instead of the phononic properties of MoSi$_2$N$_4$, which is of great significance to the operating reliability with applications in electronics.

The thermal conductivity of 2D semiconductors is one of the significant phononic properties that have attracted considerable interest \cite{zhao2020thermal,wang2017thermal}, which calls for its layered structural feature that is out-of-plane Van der Waals bond and in-plane covalent bond.
Due to the atomic thin monolayer with Van der Waals bond along the out-of-plane direction, thermal conductivities of 2D semiconductors are generally thickness-dependent in both out-of-plane, and in-plane direction \cite{yang2020synthesis}.
Several 2D semiconductors such as $h$-BN \cite{wang2016superior},phosphorene \cite{luo2015anisotropic}, and MoS$_2$ \cite{yan2014thermal} are experimentally and theoretically reported to have the thickness-dependent thermal conductivity.
As for monolayer with an in-plane covalent bond, weak covalent bonds result in lower in-plane thermal conductivity than that of graphene \cite{balandin2008superior}.
Therefore, various 2D semiconductor materials with different lattice structures and covalent bonds strength produce different thermal conductivity values.
Low thermal conductivity materials can be candidates for thermoelectric (TE) \cite{lee2016thermoelectric}, while high thermal conductivity materials can be used as thermal management materials \cite{song2018two}.
Particularly, transition metal dichalcogenides (TMDCs) with tunable band gap \cite{ramasubramaniam2011tunable} and the allotropes \cite{manzeli20172d} make possess low or high thermal conductivity.
For instance, the thermal conductivity of monolayer 1$T$-ZrSe$_2$ and 1$T$-HfSe$_2$ at room temperature is 1.2 W/m$^{-1}$K$^{-1}$ and 1.8 W/m$^{-1}$K$^{-1}$ \cite{ding2016thermoelectric}, which is benefit to thermoelectric application. In contrast, the thermal conductivity of monolayer 2$H$-MoS$_2$ is 155 W/m$^{-1}$K$^{-1}$ \cite{gu2016layer}, which can be expected to be used in thermal management applications \cite{peng2015thermal}.
Among all the 2D materials, graphene holds the highest thermal conductivity, which can be up to 3000-5300 W/m$^{-1}$K$^{-1}$ \cite{balandin2008superior,qin2018diversity}.
Such superior thermal conductivity promises its application in thermal management. However, the metallic behavior with a low on/off ratio ($<100$) at room temperature for graphene channel field-effect-transistor \cite{xia_graphene_2010}. With such consideration, TMDCs can be a candidate for both high heat conduction and excellent on/off ratio compared to graphene\cite{liu2014high,wu2013high}.

As proposed in Ref. \cite{hong2020chemical} and \cite{wang2020structure}, MoSi$_2$N$_4$ monolayer holds 1.74 eV PBE band gap and 2.30 eV HSE band gap, which is comparable to that of MoS$_2$ (1.8 eV, PBE). And the derived carrier mobility of MoSi$_2$N$_4$ is four times that of MoS$_2$. Furthermore the Young's modulus of MoSi$_2$N$_4$ is about 50 GPa, which is higher than that of MoS$_2$ by 26.8 GPa \cite{cooper2013nonlinear}. Similarity to MoS$_2$ of TMDCs, MoSi$_2$N$_4$ is also a member of MA$_2$Z$_4$ family. In addition, both MoSi$_2$N$_4$ and WSi$_2$N$_4$ have been synthesized experimentally by CVD. Geometrically, MoSi$_2$N$_4$ contains 2$H$-MoS$_2$ type MoSi$_2$ in its monolayer and it is sandwiched by two two-atomic-layer zigzag-SiN.
Given the fact that both the 2$H$-MoS$_2$ (155 W/m$^{-1}$K$^{-1}$) and Si$_3$N$_4$ (177 W/m$^{-1}$K$^{-1}$) \cite{zhou2015development} have excellent thermal conductivity, it is urgently required to study the thermal conductivity of MoSi$_2$N$_4$ and WSi$_2$N$_4$ monolayer for promoting their applications. Although machine learning\cite{MORTAZAVI2021105716} and ShengBTE\cite{Yu_2021} methods have been used to study the thermal conductivity of MoSi$_2$N$_4$ famliy, higher precision computation is always required.

In this work, we performed a systematic study of the phonon transport properties of both MoSi$_2$N$_4$ and WSi$_2$N$_4$ by solving the phonon Boltzmann transport equation (BTE) based on first-principles calculations.
Firstly, the lattice structure and phonon dispersion of the MoSi$_2$N$_4$ and WSi$_2$N$_4$ were studied. Then, the lattice thermal conductivities of them at different temperatures were calculated.
We found that the MoSi$_2$N$_4$ and WSi$_2$N$_4$ are promising thermal management materials with outstanding thermal conductivity.
The mechanism underlying the high thermal conductivity of such MoSi$_2$N$_4$-based materials is explained by analyzing the mode resolved phonon properties.
Moreover, the electronic structures were further studied to obtain a deep insight into phonon transport.
This paper systematically studied the thermal transport properties of 2D \emph{M}Si$_2$N$_4$-based materials for exploring their potential applications in thermal management and many other fields.

\section{Computational details}
\textit{Ab initio} calculations based on density functional theory (DFT) were performed using the \textit{Vienna ab initio simulation package} (VASP)\cite{kresse1993ab,kresse1996efficient}, which implements the projector augmented wave (PAW)\cite{kresse1999ultrasoft}.
Exchange-correlation energy functional is treated using the Perdew-Burke-Ernzerhof of generalized gradient approximation (GGA-PBE)\cite{perdew1996generalized}.
The wave functions are expanded in plane wave basis with a 20$\times$20$\times$1 Monkhorst-Pack\cite{monkhorst1976special} \emph{k}-sampling grid and cut-off energy of 700 eV.
A large vacuum region is set as 20 {\AA} to avoid the interactions between the monolayer and its mirrors induced by the periodic boundary conditions.
Precision of total energy convergence for the self-consistent field (SCF) calculations was as high as 10$^{-8}$ eV.
All structures are fully optimized until the  maximal Hellmann-Feynman force is less than 10$^{-8}$ eV{\AA}$^{-1}$.
To calculate the phonon dispersion, thermal conductivity, and various phonon properties, it is necessary to extract second- and third- interatomic force constants (IFCs) from first-principles calculations.
To this end, 4$\times$4$\times$1 supercells containing 112 atoms were constructed, which is sufficiently large to allow the out-of-phase tilting motion.
To extract second-order IFCs, an atom in the supercell was displaced from its equilibrium position by 0.01 $\AA$ and the Hellmann-Feynman forces were calculated based on the displaced configuration.
In order to estimate the anharmonic phonon frequencies of current systems, we extended the finite-displacement approach to prepare cubic IFCS with appropriately chosen displacement magnitude $\Delta \mu$ ($\Delta \mu$=0.04 for cubic).
Cut-off radii of 18a0 and 14a0 (where a0 is Bohr
radius) were used for harmonic and cubic IFCs, respectively.
The harmonic IFCs with all possible harmonic terms between each two atoms will be included, resulting in 19 displacement configurations.
The cubic IFCs within a cut-off radius of 14a0 result in 2771 displacement configurations.
The convergence of the dependence of cut-off radius on
phonon-band dispersion and mode Gr\"{u}neisen parameters was
checked.
Lattice thermal conductivity ($\kappa_L$) and relative phonon properties were determined by solving the phonon BTE, as implemented in the ALAMODE~\cite{tadano2014anharmonic} package. 
The $\kappa_L$ is estimated by the BTE within RTA through the following equation:
\begin{equation}
    \kappa_L^{\alpha\beta}(T)=\frac{1}{NV}\sum_qC_q(T)\nu_q^{\alpha}(T)\nu_q^{\beta}(T)\tau_q(T),
\end{equation}
where $V$, $C_q(T)$, $\nu_q(T)$, and $\tau_q(T)$ are the unit cell volume, mode specific heat, phonon group velocity, and phonon lifetime, respectively.

\begin{figure}
\centering
  \includegraphics[width=15cm]{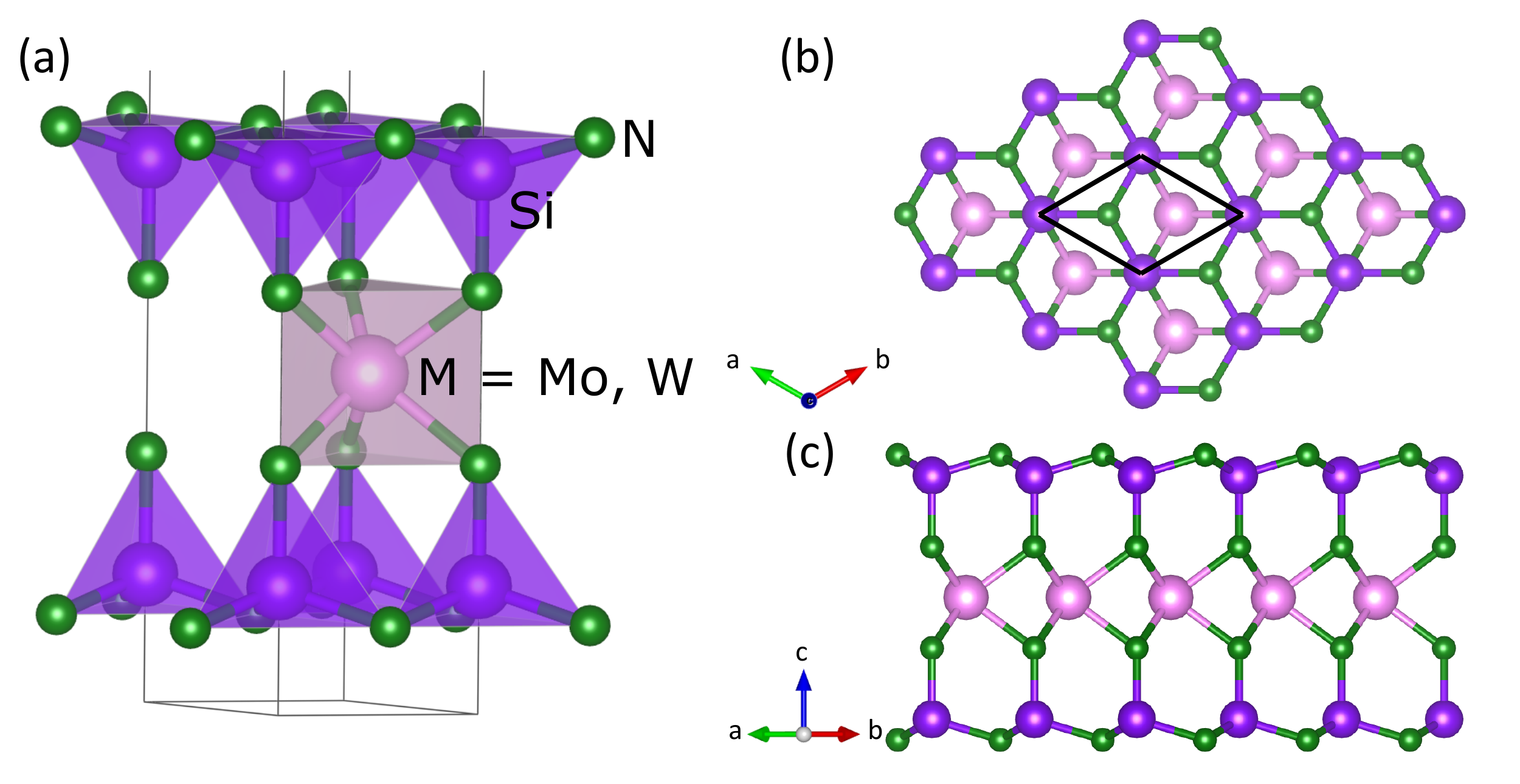}
  \caption{The top and side views of structures of MoSi$_2$N$_4$ and WSi$_2$N$_4$.}
  \label{fgr:structure}
\end{figure}

\section{Results and discussion}
\subsection{Lattice structures of monolayer MoSi$_2$N$_4$ and WSi$_2$N$_4$}

MoSi$_2$N$_4$ and WSi$_2$N$_4$ are new septuple-atomic-layer monolayer materials reported in Ref.\cite{hong2020chemical}, as shown in Fig. \ref{fgr:structure}.
This monolayer is built up by septuple atomic layers of N-Si-N-\emph{M}-N-Si-N (\emph{M} = Mo, W), which can be structurally constructed by inserting 2$H$-MoS$_2$-type \emph{M}N$_2$ into $\alpha$-InSe-type SiN.
Interestingly, the space group of MoSi$_2$N$_4$ (P\={6}m2, No. 187) is consistent with that of 2$H$-MoS$_2$ and $\alpha$-InSe monolayer.
In detail, in Fig. \ref{fgr:structure}(a), Si locates in the center of a tetrahedron formed by four N atoms, and Mo is in a triangular prism consisted of six N atoms. Notably, MoSi$_2$N$_4$ described above is identified as the most energetically favorable structure among thirty structures proposed in Ref.\cite{wang2020structure} in combination with first-principles structural optimization (lattice parameters and space group are listed in Table \ref{tbl:structure}).

\begin{table}[htp]
\small
  \caption{\ Lattice parameters of monolayer MoSi$_2$N$_4$ and WSi$_2$N$_4$ (unit {\AA}).}
  \label{tbl:structure}
  \begin{tabular*}{\textwidth}{@{\extracolsep{\fill}}llllc}
    \hline
    Compound & Space group & a & Thickness  \\
    \hline
    MoSi$_2$N$_4$ & P\={6}m2 & 2.91 & 6.79 &   \\
    WSi$_2$N$_4$ & P\={6}m2 & 2.91 & 7.02 & \\
    \hline
  \end{tabular*}
\end{table}

\subsection{Phonon dispersion and density of states}
\begin{figure*}[htp]
 \centering
 \includegraphics[width=15cm]{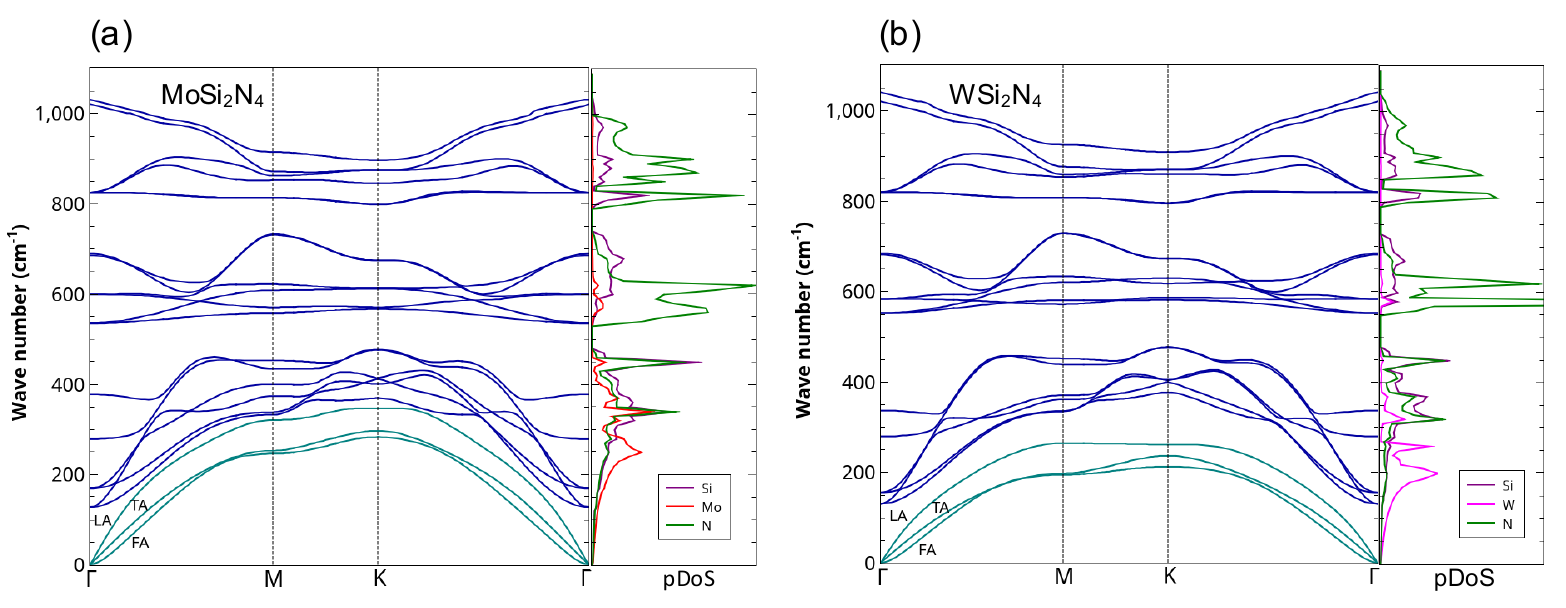}
 \caption{The phonon dispersions of MoSi$_2$N$_4$ and WSi$_2$N$_4$.}
 \label{fgr:structure}
\end{figure*}

To study the phonon transport properties of monolayers MoSi$_2$N$_4$ and WSi$_2$N$_4$, phonon dispersion calculations are firstly performed based on the finite displacement difference method.
The phonon bands are shown in Fig.\ref{fgr:structure}, it is noted that no imaginary part existing in monolayers MoSi$_2$N$_4$ and WSi$_2$N$_4$, indicating the dynamic stability of the two monolayer compounds.
The primitive cell of monolayer MoSi$_2$N$_4$(WSi$_2$N$_4$) has seven atoms. Thus there are three acoustic phonon branches and eighteen optical branches.
The three lowest phonon branches are acoustic phonon branches, \textit{i.e.} the out-of-plane flexural acoustic (FA) branch, the in-plane transverse acoustic (TA) branch, and the in-plane longitudinal acoustic (LA) branch,  present linear behavior when approaching the $\Gamma$ point, while the flexural acoustic (FA) phonon branch shows a quadratic behavior, which is consistent with our previous results\cite{qin2017anomalously}.
Similar behaviors are always found in 2D materials, and the consistency ensures the accuracy of the obtained thermal conductivity.

In addition, the two monolayer compounds show very similar dispersion curves along the path passing through the main high symmetry K-points in the irreducible Brillouin zone (IBZ).
The phonon dispersions of monolayer MoSi$_2$N$_4$ and WSi$_2$N$_4$ are separated into three regions. There exists a gap between the regions. The gap between these two compounds is also quite similar.
Interestingly, the frequencies of acoustic branches of MoSi$_2$N$_4$ and WSi$_2$N$_4$ are pretty similar and comparable with the common 2D materials, such as WS$_2$ and MoS$_2$.
The high frequency of acoustic branches indicates that the phonon harmonic vibrations of MoSi$_2$N$_4$ and WSi$_2$N$_4$ are strong, which will significantly affect the phonon transport properties.
However, as shown in Fig. \ref{fgr:structure}, TA, FA, and LA branches (highlighted as a green line) of MoSi$_2$N$_4$ and WSi$_2$N$_4$ are different. The LA branch of MoSi$_2$N$_4$ interacts with other optical modes, and the frequency is larger than that of WSi$_2$N$_4$. This phenomenon will also impact the thermal transport properties; we will discuss them in the next section.

As revealed by the partial density of states (pDOS), high-frequency optical phonon branches above the gap are mainly contributed
by the vibration of the light N atom.
In the low-frequency range around 200 cm$^{-1}$, the contribution of W atoms and Mo atoms to the DOS of the MoSi$_2$N$_4$ and WSi$_2$N$_4$ are the main parts.
In frequency range around 400 cm$^{-1}$, partial DOS of Si atoms has the same level with N atoms, while above the frequency of 400 cm$^{-1}$, partial DOS of N atoms is more significant than that of other atoms, and the transition metals Mo and W contribute rarely.

\subsection{Phonon transport properties}
\begin{figure}[h!]
\centering
  \includegraphics[width=15cm]{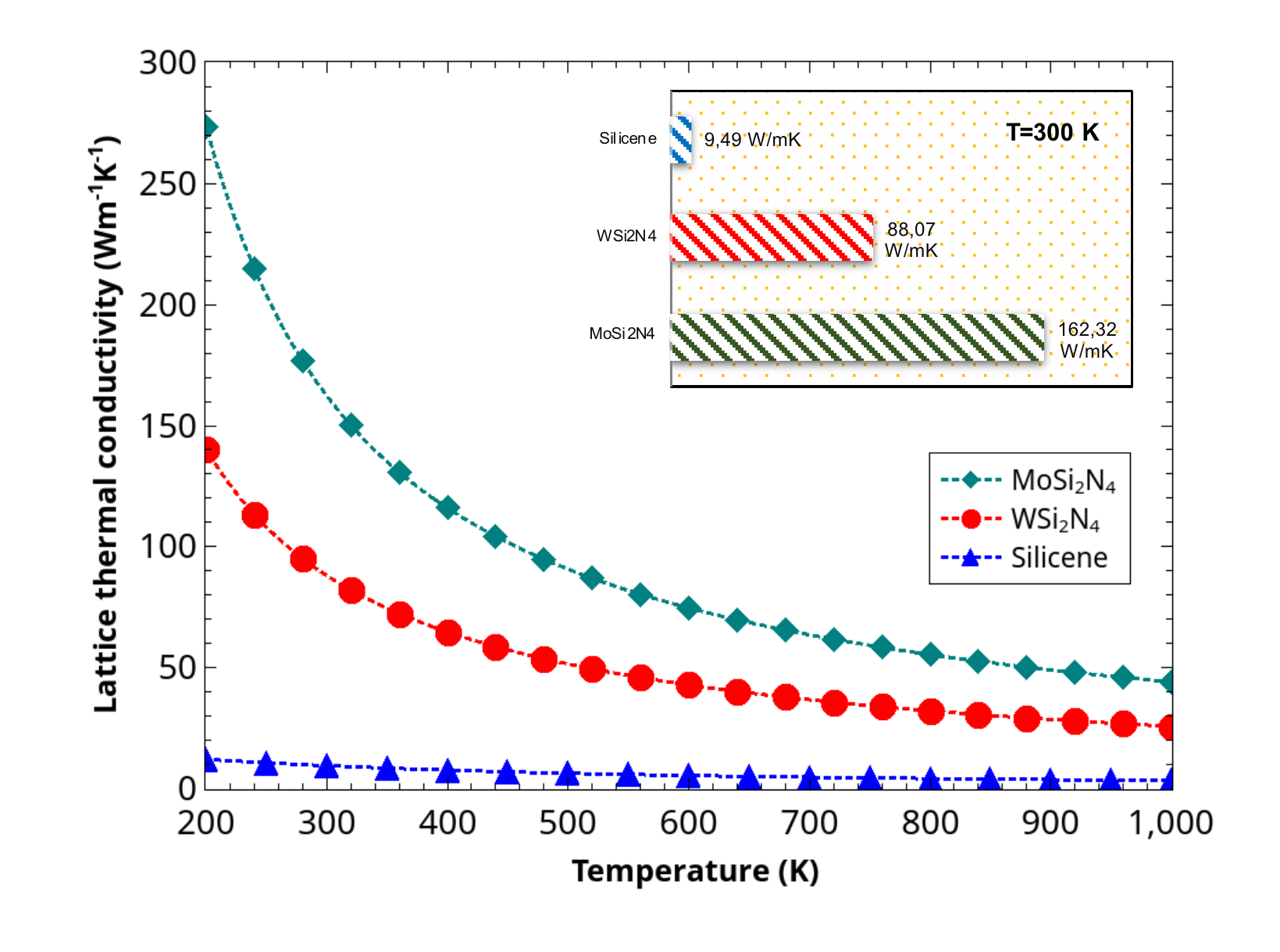}
  \caption{Temperature (200-1000 K) dependent thermal conductivities of monolayer MoSi$_2$N$_4$ and WSi$_2$N$_4$.}
  \label{fgr:LTC}
\end{figure}

Based on the harmonic and anharmonic IFCs, the lattice thermal conductivities of monolayer MoSi$_2$N$_4$ and WSi$_2$N$_4$ are calculated solving the BTE.
In Fig. \ref{fgr:LTC}, thermal conductivities of the two systems at different temperatures with silicene as reference for comparison are presented.
As one can see that, the intrinsic lattice thermal conductivities are temperature dependent, which is approximately proportional to the inverse temperature of 1/T, consistent with the expected behavior of crystalline materials in both bulk and 2D forms.
It is noted that MoSi$_2$N$_4$ has larger thermal conductivities.
In contrast, the thermal conductivities of WSi$_2$N$_4$ are lower, which may be owing to the low frequencies of TA, FA, and LA branches.
Comparing with the common excellent thermoelectric materials, such as ZrSe$_2$ \cite{ding2016thermoelectric} and HfSe$_2$ \cite{gu2016layer}, the thermal conductivities of MoSi$_2$N$_4$ and WSi$_2$N$_4$ are huge, 162 W/mK and 88 W/mK at room temperature respectively.
Such large thermal conductivities limit the MoSi$_2$N$_4$ and WSi$_2$N$_4$ being the promising thermoelectric material.
However, these two compounds with such larger thermal conductivities present the promising potential for thermal management materials.
The MoSi$_2$N$_4$ and WSi$_2$N$_4$ possess rather high thermal conductivity compared to a lot of thermal management materials, such as MoS$_2$, silicene, phosphorene, \textit{etc}.\cite{luo2015anisotropic}
In this work, the thermal conductivity of silicene as the comparison is plotted in Fig. \ref{fgr:LTC}.
As one can see, the intrinsic lattice thermal conductivities of MoSi$_2$N$_4$ and WSi$_2$N$_4$ are 7 and 4 times the one for silicene.

\begin{figure}[h!]
\centering
  \includegraphics[width=13cm]{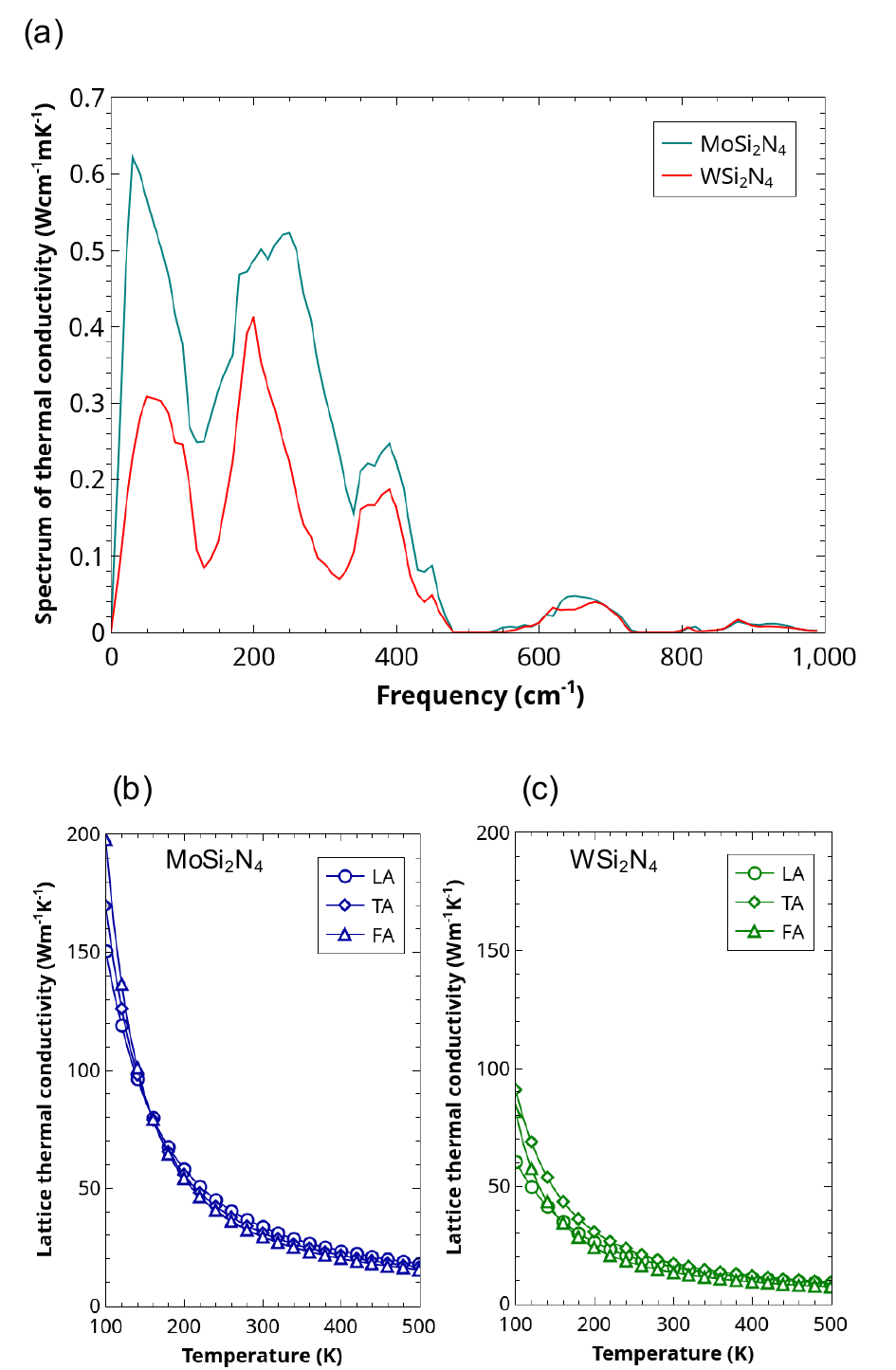}
  \caption{ (a) The spectrum of thermal conductivities (at 300 K) and (b,c) acoustic phonon branch contributions for overall thermal conductivities of monolayer MoSi$_2$N$_4$ and WSi$_2$N$_4$.}
  \label{fgr:LTC-mode}
\end{figure}

%%%%%
To understand the underlying mechanism responsible for
the mode contributed thermal conductivity of monolayers MoSi$_2$N$_4$ and WSi$_2$N$_4$. We plot the spectrum of thermal conductivities and absolute contribution from each acoustic phonon branch [FA, TA, and LA as marked in Fig. \ref{fgr:LTC-mode}] to the overall thermal conductivity in Fig. \ref{fgr:LTC}.
As shown in Fig. \ref{fgr:LTC-mode}(a), the low-frequency phonons (below a frequency of 400 cm$^{-1}$) of MoSi$_2$N$_4$ and WSi$_2$N$_4$ dominate the thermal conductivity contributions. Especially, the contribution from FA, TA, and LA phonon branches (below a frequency of 200 cm$^{-1}$) contribute the most proportion of thermal conductivities of both of them.
Further,  the high-frequency phonons of MoSi$_2$N$_4$ and WSi$_2$N$_4$ hardly contribute as heat carriers.
In Fig. \ref{fgr:LTC-mode}(b and c), we compared each acoustic phonon branch (FA, TA, and LA as marked in Fig. \ref{fgr:LTC-mode}) to the overall thermal conductivity. Each acoustic phonon branch contributes quite similarly to both materials. However, the FA, TA, and LA phonon branches of MoSi$_2$N$_4$ are almost two times larger than the values of WSi$_2$N$_4$; the results are also consistent with the deduction from phonon dispersion of them.
The main contribution of phonons in the low-frequency region of two-dimensional materials to the thermal conductivity has been confirmed in many materials.\cite{liu2016disparate}

\begin{figure}[h!]
\centering
  \includegraphics[width=15cm]{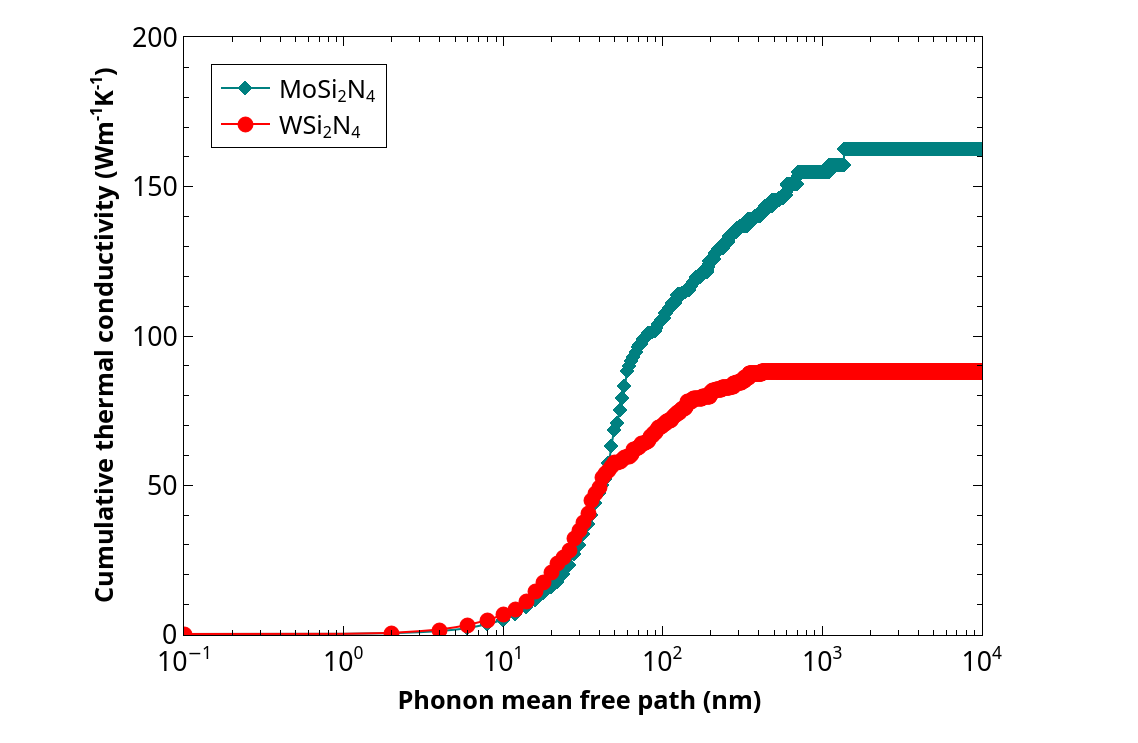}
  \caption{ Comparison of the cumulative lattice thermal conductivities of monolayer MoSi$_2$N$_4$ and WSi$_2$N$_4$ with respect to phonon mean free path (MFP) at 300 K.}
  \label{fgr:CLTC}
\end{figure}

%%%%%
Finally, to understand the grain-size effect quantitatively, the cumulative lattice thermal conductivity was analyzed concerning the mean free path (MFP) of phonons.
The cumulative thermal conductivities with respect to MFP for the MoSi$_2$N$_4$ and WSi$_2$N$_4$ are plotted in Fig. \ref{fgr:CLTC}.
The MFPs corresponding to 50\% of the cumulative lattice thermal conductivities for MoSi$_2$N$_4$ and WSi$_2$N$_4$ at 300 K are 80 and 50 nm, respectively.
The MFP helps study the size effect on the ballistic vs. diffusive phonon transport. This quantity is crucial for the thermal design to modulate the thermal conductivity in the small-grain limit.

\subsection{Mode level analysis}
\begin{figure}[h!]
\centering
  \includegraphics[width=13cm]{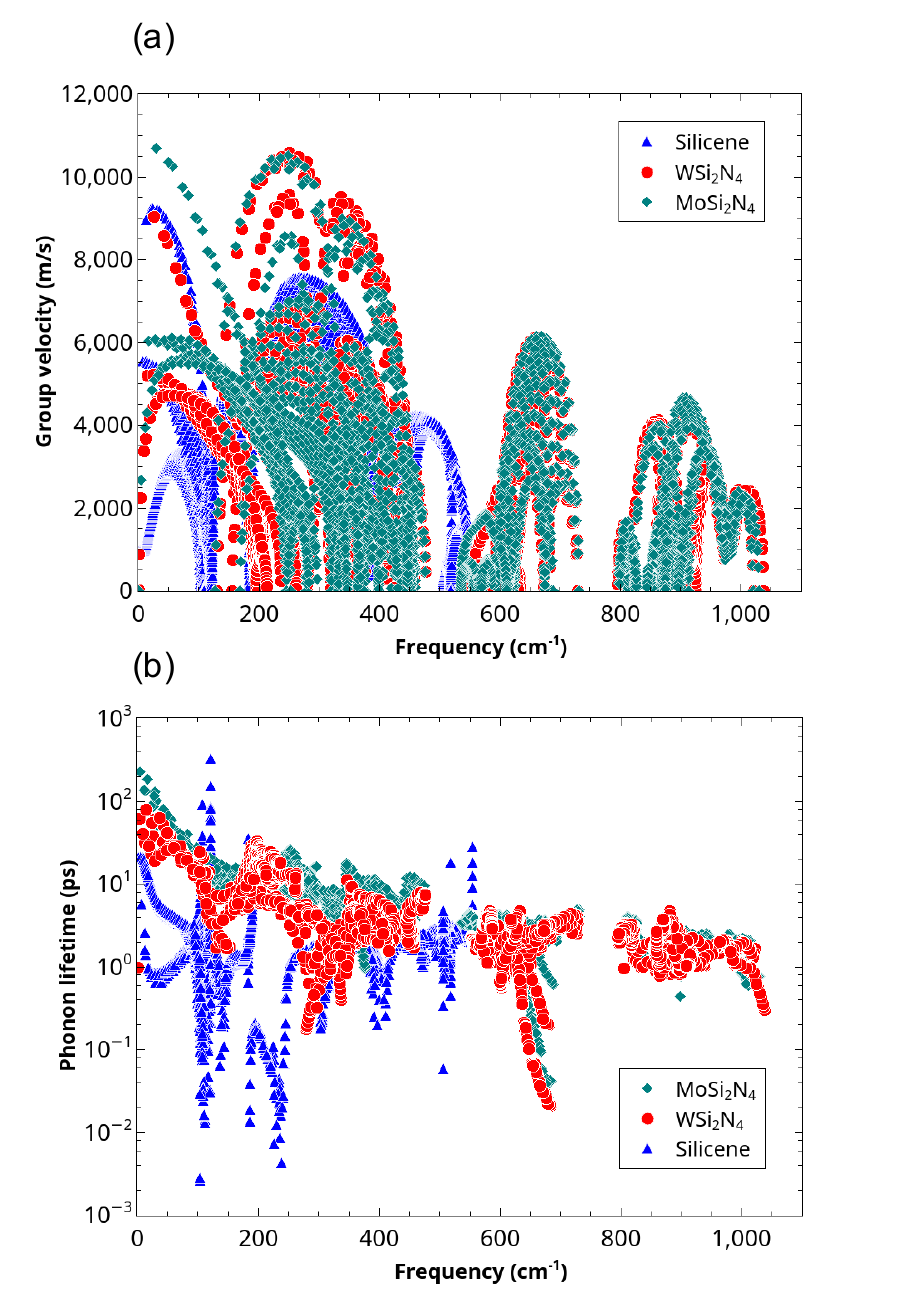}
  \caption{The comparison of mode level (a) phonon group velocity and (b) phonon lifetime of monolayer MoSi$_2$N$_4$, WSi$_2$N$_4$ and silicene at 300K.}
  \label{fgr:mode-level}
\end{figure}
%%%%%phonon velocity
To gain insight into the phonon transport in MoSi$_2$N$_4$ and WSi$_2$N$_4$, we performed a detailed analysis on the mode level phonon group velocity and lifetime (relaxation time).
Comparison of the mode level phonon group velocity of MoSi$_2$N$_4$, WSi$_2$N$_4$ and silicene as a function of frequency at 300 K are shown in Fig.\ref{fgr:mode-level}(a).
It is worth noted that the overall phonon group velocity of monolayer MoSi$_2$N$_4$, WSi$_2$N$_4$ are on the same order of magnitude, which is larger than that of monolayer silicene.
As for these three materials, it is interesting to notice that the acoustic phonon branches have large phonon group velocities.
Furthermore, the high-frequency phonon branches (above a frequency of 400 cm$^{-1}$) for MoSi$_2$N$_4$, WSi$_2$N$_4$ have also relatively large phonon group velocities, which is distinctly different from silicene.
Besides, the phonon velocity of acoustic branches of MoSi$_2$N$_4$ is larger than the values of WSi$_2$N$_4$, and the rest of the branches are almost identical.
It is also confirmed that TA, FA, and LA manipulate thermal conductivity between MoSi$_2$N$_4$ and WSi$_2$N$_4$. Finally, to understand the grain-size effect quantitatively, the cumulative lattice thermal conductivity was analyzed concerning the mean free path (MFP) of phonons.
The cumulative thermal conductivities with respect to MFP for the MoSi$_2$N$_4$ and WSi$_2$N$_4$ are plotted in Fig. \ref{fgr:CLTC}.
The MFPs corresponding to 50\% of the cumulative lattice thermal conductivities for MoSi$_2$N$_4$ and WSi$_2$N$_4$ at 300 K are 80 and 50 nm, respectively.
The MFP helps study the size effect on the ballistic vs. diffusive phonon transport. This quantity is crucial for the thermal design to modulate the thermal conductivity in the small-grain limit.

%%%%Lifetime
In addition, the phonon lifetimes of MoSi$_2$N$_4$, WSi$_2$N$_4$ and silicene at 300 K are plotted in the Fig. \ref{fgr:mode-level}(b).
It can be seen that the overall lifetime of
phonon branches of silicene are smaller than that of MoSi$_2$N$_4$, WSi$_2$N$_4$, which might be due to the enhanced phonon-phonon scattering in silicene, so it has a lower lattice thermal conductivity.
Considering the same magnitude of group velocity as MoSi$_2$N$_4$,
WSi$_2$N$_4$ possesses lower thermal conductivity due to its small phonon lifetime.

\subsection{Phonon anharmonicity}
\begin{figure}[h!]
\centering
  \includegraphics[width=15cm]{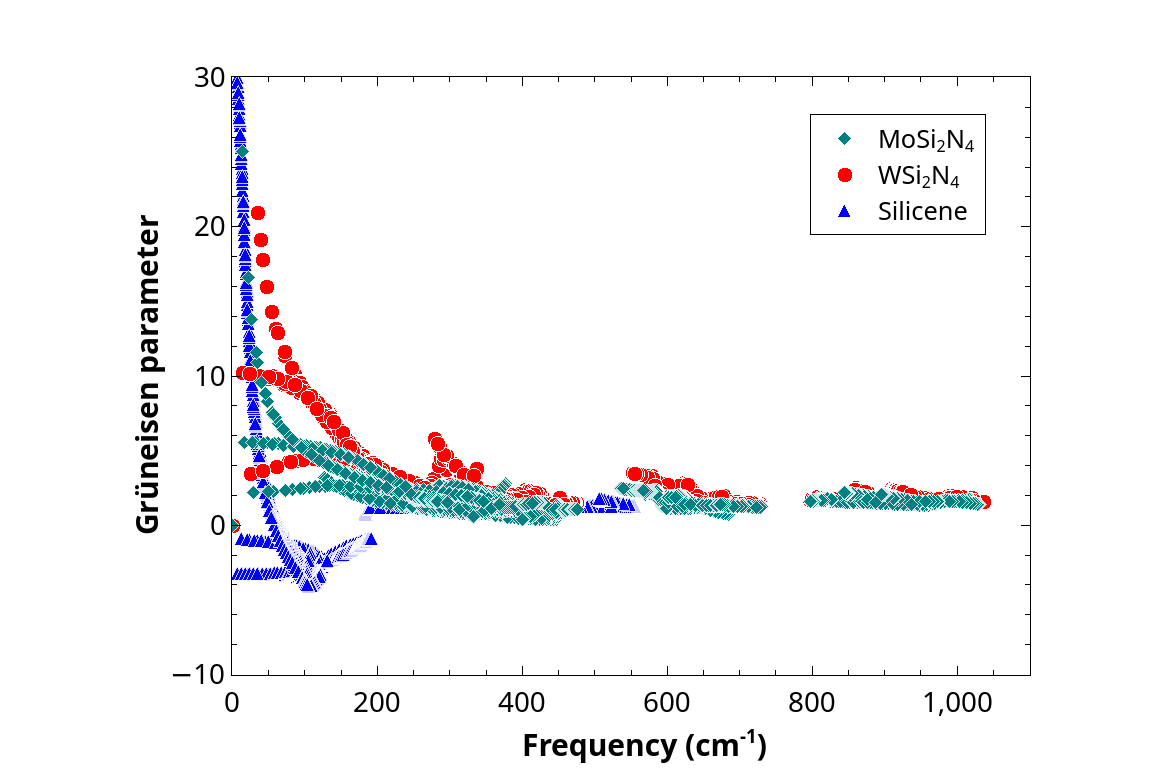}
  \caption{The mode level Gr\"{u}neisen parameters for MoSi$_2$N$_4$, WSi$_2$N$_4$ and silicene.}
  \label{fgr:grun}
\end{figure}

It is well known that the phonon-phonon scattering process is
determined by the anharmonic nature of structures, whose magnitude can be roughly quantified by the Gr\"{u}neisen parameter. To this end, we examine the phonon anharmonicity of these materials by calculating the Gr\"{u}neisen parameter.
As shown in Fig. \ref{fgr:grun}, the magnitude of Gr\"{u}neisen parameter for silicene is obviously larger than MoSi$_2$N$_4$ and WSi$_2$N$_4$, meaning stronger phonon anharmonicity in silicene.
The strong phonon-phonon scattering due to the anharmonicity leads to the small phonon lifetime of silicene (Fig. \ref{fgr:mode-level}(b)), and thus leads to the low thermal conductivity of silicene (Fig.\ref{fgr:LTC}).
For the same reason, the Gr\"{u}neisen parameter of MoSi$_2$N$_4$ smaller than that of WSi$_2$N$_4$, which is the underlying reason for the small phonon lifetime and further lower thermal conductivity of WSi$_2$N$_4$ than MoSi$_2$N$_4$.
Significantly, the deviation of the Gr\"{u}neisen parameter between MoSi$_2$N$_4$ and WSi$_2$N$_4$ mainly occurs at low-frequency range (below a frequency of 200 cm$^{-1}$), indicating the acoustic phonon branches dominate the thermal transport properties.

\subsection{Insight from electronic structures}
\begin{figure*}[htp]
\centering
  \includegraphics[width=13cm]{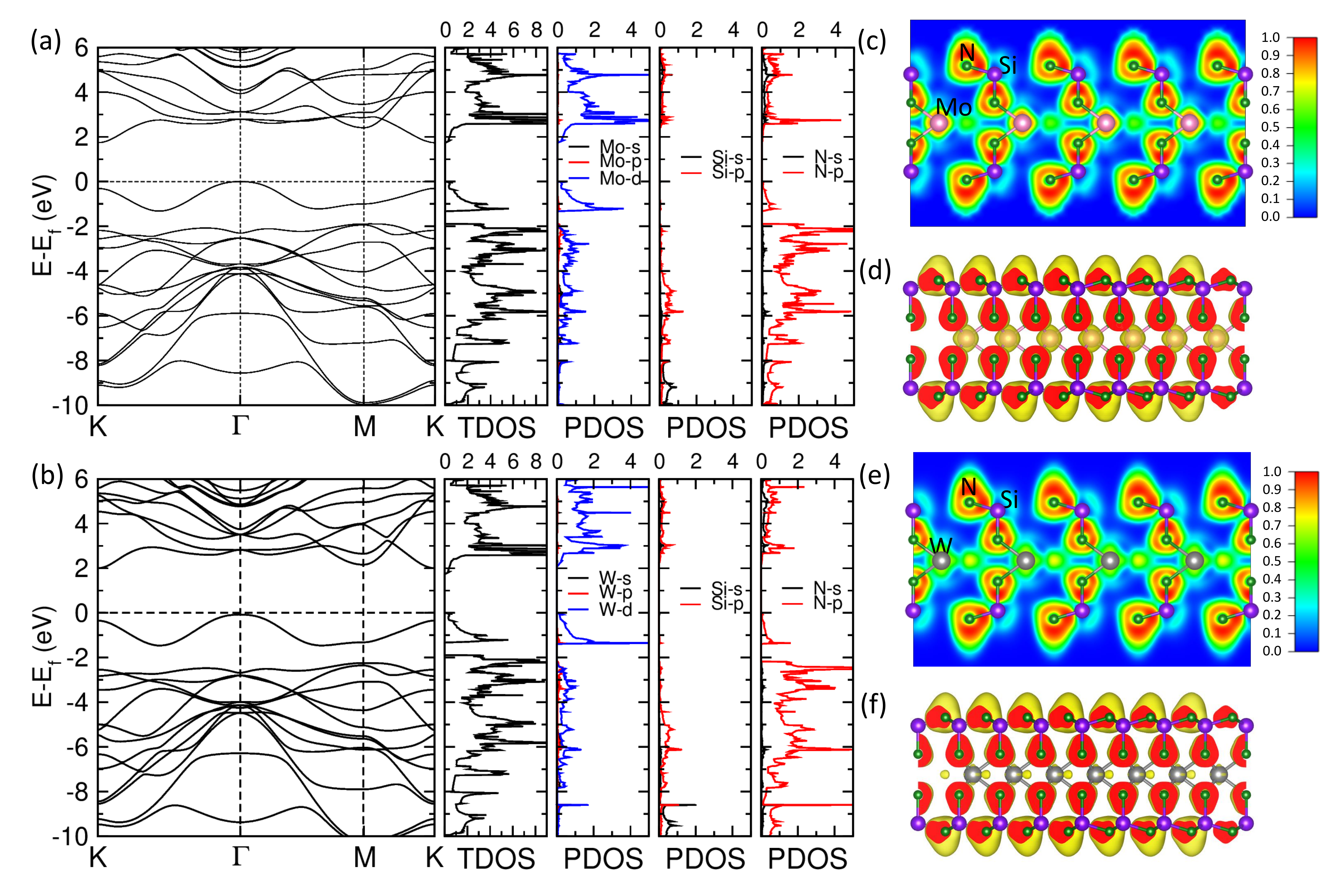}
  \caption{(a,b) The electronic structures for MoSi$_2$N$_4$, WSi$_2$N$_4$, (c,e) 2D ELF images for MoSi$_2$N$_4$, WSi$_2$N$_4$, and (d,f) 3D isosurface images with ELF = 0.6 for MoSi$_2$N$_4$, WSi$_2$N$_4$.}
  \label{fgr:elec}
\end{figure*}

To understand the fundamental mechanism underlying the phonon thermal transport, the electronic structures of MoSi$_2$N$_4$ and WSi$_2$N$_4$ monolayers are derived by first-principles calculations, as comparably shown in Fig. \ref{fgr:elec}(a and b).
MoSi$_2$N$_4$ and WSi$_2$N$_4$ are semiconductors with indirect band gap about 1.74 and 2.08 eV, which are also confirmed by the density of states (DOS) as shown in Fig. \ref{fgr:elec}(c,e). For both MoSi$_2$N$_4$ and WSi$_2$N$_4$ monolayers, we find that the valence bands from -10 to -2 eV mainly originate from $p$-orbits of N. In contrast, the bands from -1.5 to 6 eV are mainly dominated by the $d$-orbits of \emph{M} (\emph{M} = Mo, W) and are weakly contributed by the $p$-orbits of N. Such fact implies strong orbital hybridization between N-$p$ and \emph{M}-$d$ orbits. To further explore the bonding characteristics of \emph{M}Si$_2$N$_4$ monolayer, the electron localization function (ELF) is plotted in Fig. \ref{fgr:elec}(c,e) ranging from 0 (blue) to 1 (red). ELF = 1 means perfect localization, while ELF = 0.5 is the probability of electron-gas-like pair. In \emph{M}Si$_2$N$_4$ monolayers, the electrons are largely localized around N atoms and decayed from \emph{M} and Si atoms, indicating that the electrons are transferred from \emph{M} and Si to N to form bonds with ionic characteristics between \emph{M}, Si and N. Furthermore, compared with Mo atom, smaller ELF value near W atoms indicates stronger ionic bonding of W-N than Mo-N, which is consistent with the TMDCs\cite{peng_thermal_2016,huang_roles_2015}. A close inspection of the 2D ELF images finds that there is an area between two W atoms in WSi$_2$N$_4$ with larger ELF values than that of MoSi$_2$N$_4$, which is further verified by the 3D isosurface images (ELF = 0.6) as shown in Fig. \ref{fgr:elec}(d,f). The form of this 'lone pair electrons liked' area could be responsible for the lower thermal conductivity of WSi$_2$N$_4$\cite{nielsen_lone_2013}.

Generally, strong interatomic bonding and low average atomic mass are benefit to high thermal conductivity of 2D materials\cite{SLACK1973321}. Therefore, the Young's modulus and average atomic mass of \emph{M}Si$_2$N$_4$ are derived in Table~\ref{tbl:mechanic}. For more detail, in the linear elastic regime the Young's modulus of 2D materials (Y$^{2D}$) is possible to estimate by the elastic constants C11 and C12, the form of Y$^{2D}$ as follows\cite{politano_probing_2015,zhang_superior_2018}
\begin{equation}\label{Y2D}
  Y^{2 D}=\frac{C_{11}^{2}-C_{12}^{2}}{C_{11}}
\end{equation}
Note that, the elastic constant is rescaled by $d_0/h$ to obtain the effective elastic constant, where $d_0$ is slab model length along thickness direction and $h$ is the thickness of monolayer. As listed in Table~ref{tbl:mechanic}, we can find that the Young's modulus of \emph{M}Si$_2$N$_4$ is more than twice than that of MoS$_2$ and silicene, which indicates stronger interatomic bonding in \emph{M}Si$_2$N$_4$. Furthermore, the average atomic mass ($\bar{M}$ = $M_t/n$, where $M_t$ is total mass per formula cell, and $n$ is the number of atoms per formula cell) of MoSi$_2$N$_4$ is about half of MoS$_2$ and comparable to that of silicene, which may induce it high thermal conductivity.

\begin{table}[h!]
\small
  \caption{\ The average atomic mass ($\bar{M}$), Young's modulus (Y$^{2D}$, GPa) and thickness ($h$, \AA) of MSi$_2$N$_4$, MoS$_2$ and silicene.}
  \label{tbl:mechanic}
  \begin{tabular*}{\textwidth}{@{\extracolsep{\fill}}lccc}
    \hline
    Compound & $\bar{M}$ & Y$^{2D}$ & $h$    \\     \hline
    MoSi$_2$N$_4$ & 29.74 & 474.75 & 10.28 \\
    WSi$_2$N$_4$  & 42.29 & 499.83 & 10.29 \\
    MoS$_2$       & 53.36 & 205.81 & 6.04 \\
    silicene      & 28.09 & 207.94 & 2.97 \\
    \hline
  \end{tabular*}
\end{table}

\section{Conclusions}
In summary, by solving the phonon BTE based on the first-principles calculations, we have performed a comprehensive study on the phonon transport properties of MoSi$_2$N$_4$ and WSi$_2$N$_4$ and made a thorough comparison with silicene. The thermal conductivities of MoSi$_2$N$_4$ and WSi$_2$N$_4$ are found to be 162 W/mK and 88 W/mK at room temperature respectively, which are 7 and 4 times the one for silicene.
These results show that, the MoSi$_2$N$_4$ and WSi$_2$N$_4$ have promising potential being thermal management materials.
To understand the underlying mechanism for the high thermal conductivity of MoSi$_2$N$_4$ and WSi$_2$N$_4$, the systematic analysis is performed based on the study of contribution from each phonon branch and comparison among the mode level phonon group velocity and lifetime. The root reason for the high thermal conductivity of MoSi$_2$N$_4$ and WSi$_2$N$_4$ is that the high group velocity of these two materials. The phonon Gr\"{u}neisen parameter is further analyzed to understand the phonon-phonon scattering. And the Gr\"{u}neisen parameter of MoSi$_2$N$_4$ smaller than that of WSi$_2$N$_4$, which is the underlying reason for the small phonon lifetime and further lower thermal conductivity of WSi$_2$N$_4$ than MoSi$_2$N$_4$.
Therefore, our study offers fundamental understanding of thermal transport properties in monolayer MoSi$_2$N$_4$ and WSi$_2$N$_4$ within the framework of BTE and the electronic structures from the bottom, which will enrich the studies and exploring of novel \emph{M}Si$_2$N$_4$ type two dimensional thermal management materials.

\section*{Conflicts of interest}
There are no conflicts to declare.

\section*{Acknowledgements}
The authors gratefully thank Manuel Richter for providing the SOC strength data and helpful discussions.
This work was also supported by the Deutsche Forschungsgemeinschaft - Project-ID~405553726 - TRR 270.
X.-Q.C. is supported by the National Science Fund for Distinguished Young Scholars (Grant No. 51725103).
G.Q. is supported by the Fundamental Research Funds for the Central Universities (Grant No. 531118010471), the National Natural Science Foundation of China (Grant No. 52006057), and the Changsha Municipal Natural Science Foundation (Grant No. kq2014034).
The Lichtenberg high-performance computer of the TU Darmstadt is gratefully acknowledged for the computational resources where the calculations were conducted for this project.

\bibliography{achemso-demo}

\end{document}